\documentclass[12pt]{iopart}

\usepackage{graphicx}
\usepackage{subfig}
\usepackage{iopams}
\usepackage[abbr]{harvard}

\begin{document}

\jl{12}

\title[Number of beams in IMRT, Single-Arc]{The number of beams in IMRT - theoretical investigations and implications for single-arc IMRT}

\author{Thomas Bortfeld}
\address{Massachusetts General Hospital and Harvard Medical School, Department of Radiation Oncology, 30 Fruit St, Boston, MA 02114, USA}
\ead{tbortfeld@partners.org}

\begin{abstract}
\noindent
The first purpose of this paper is to shed some new light on the old
question of selecting the number of beams in intensity-modulated
radiation therapy  (IMRT). The second purpose is to illuminate the related issue of
discrete static beam angles vs. rotational techniques, which has
recently re-surfaced due to the advancement of volumetric arc therapy
(VMAT). A specific objective is to find {\em analytical}
expressions that allow one to address the points raised above. To make the problem
mathematically tractable, it is assumed that the depth dose is flat and
that the lateral dose profile can be approximated by polynomials,
specifically Chebyshev polynomials of the first kind, of finite
degree. The application of methods known from image reconstruction
then allows one to answer the first question above as follows: The
required number of beams is determined by the maximum degree of the
polynomials used in the approximation of the beam profiles, which is a
measure of the dose variability.   There is nothing to be gained by
using more beams. In realistic cases, in which the variability of the
lateral dose profile is restricted in several ways, the required number
of beams is of the order of 10 to 20. The consequence of delivering
the beams with a `leaf sweep' technique during continuous rotation of
the gantry,  as in VMAT, is also derived in analytical form. The main
effect is that the beams fan out, but the effect near the axis of
rotation is small.  This result can serve as a theoretical
justification of VMAT. Overall the analytical derivations in this
paper, albeit based on strong simplifications, provide new insights into,
and a deeper understanding of, the beam angle problem in IMRT. The
decomposition of the beam profiles into well-behaved and easily
deliverable smooth functions, such as Chebyshev polynomials, could be
of general interest in IMRT treatment planning.
\end{abstract}
{\it Keywords\/}: IMRT, number of beams, Chebyshev polynomial, single-arc, VMAT
\maketitle

\section{Introduction}
The problem of selecting the number and direction of beams in intensity-modulated
radiation therapy (IMRT) has been investigated by many and has been the subject of
debate as early as in 1995 \cite{Mohan1995}. One may argue that even today, after
many more papers have been published on this topic, the results are not entirely convincing,
and not practically useful overall. In fact, in today's clinical practice, the number and
direction of beams often has to be found by trial and error. Two `schools' of IMRT delivery
co-exist today: in the more common approach, a relatively small number
(typically of the order of 10 or less) of fixed intensity-modulated beams is
delivered with a multileaf collimator (MLC). In another approach, a very large
number ($>50$) of beams is used in the rotational tomotherapy
approach \cite{Mackie1993,Carol1994}. More recently, another rotational
IMRT approach, `Single-Arc', has been developed \cite{Cameron2005,Ulrich2007,Otto2008,Wang2008}
and has found a lot of interest as a commercial product (RapidArc, VMAT). In the Single-Arc
approach there is no intensity modulation for any one beam angle, but the radiation field
shape is varied dynamically and rapidly with an MLC as the gantry rotates around  the patient.
In the rotational approaches, the selection of beam angles is not a problem, but the
distribution of dose over large volumes of healthy tissues has been a concern.

Intuitively one would expect that using more beams will always help to shape radiation
dose distributions to match the tumor target volume, even though it is clear that there is a
point of diminishing return. In this paper I will address the `how many beams' problem from a
theoretical viewpoint. I will introduce strong simplifications to make the problem
mathematically tractable. The mathematical approach used here is similar to the one from
an earlier paper on image reconstruction \cite{Bortfeld1999}. One of the results
is that, contrary to the intuition,  there is no benefit whatsoever in increasing
the number of beams beyond a certain threshold, which depends on the achievable
amount of intensity modulation per beam. I will present some illustrative examples.

In the last methodological chapter I will take the step from standard IMRT with a
small number ($K$) of fixed beam angles towards rotational therapy.
From what I said above it is clear that dynamically rotating the beam
cannot yield better dose distributions, but it can potentially reduce the treatment time.
First I will derive a simple equation to describe the effect of rotating the gantry
while IMRT is being delivered with a sweep technique \cite{Convery1992}. I will apply this
equation to the delivery of polynomial beam profiles over short arcs  of $180^\circ / K$.
As a result, the impact of this small angle rotation is small, thus serving as a
theoretical justification of the Single-Arc approach.

\section{Polynomial approximations}
Let us adopt the coordinate system from an earlier paper \cite{Bortfeld1999}.
Here the $x$-axis points from left to right, and the $y$ axis points to the radiation
source at the $\phi=0^\circ$ gantry position. In other words,
at $\phi=0^\circ$ gantry angle, the beam goes in the $-y$ direction, i.e., downwards.
The gantry angle is counted positive in the {\em counter-clockwise} direction; note,
however, that the gantry angle was defined in reversed (i.e., in clockwise) orientation
in a more recent paper \cite{Bortfeld2009}.

I will use a zero-order approximation of the depth dose distribution, i.e., I assume that
the depth-dose
is flat. This means in particular that I ignore beam
divergence and the variability of scatter as a function of depth. Incidentally,
one can consider this approximation as a first order approximation (linear depth-dose profile)
if one uses parallel-opposed beams, but parallel-opposed beams will not be considered further
 in the rest of this paper. Within this approximation one can now write the two-dimensional
dose resulting from a beam impinging under an angle $\phi$ with fluence (intensity)
profile $f_\phi(p)$ as
\begin{equation} \label{EQ_BACKPROJ}
D_\phi(\vec r) = f_\phi(\vec r\cdot \vec n_\phi).
\end{equation}
Here $\vec r=(x,y)$ and $\vec n_\phi = (\cos\phi,\sin\phi)$ is the unit
vector perpendicular to the beam direction. In image reconstruction terminology,
equation (\ref{EQ_BACKPROJ}) is a `backprojection', i.e., a `smearing out' of
fluence values along the rays of the beam \cite{Bortfeld1999}. In section \ref{SEC_DIS}
I will discuss why I believe that, in spite of this strong simplification, the
results are not only of theoretical but also of some practical relevance.

Let us now make one additional assumption, namely that one can write the fluence (intensity)
profiles as a set of polynomials of degree $m$. This should always be possible because in
practice the fluence profile is well behaved thanks to the finite source size, scatter,
partial transmission through the leaf ends of the MLC, among other reasons.

Under these assumptions, it follows from an earlier paper \cite{Bortfeld1999} that  one can
produce any dose distribution of the form
\begin{equation} \label{EQ_APP2D}
D(\vec r) = \sum\limits_{m=0}^{M-1} \int\limits_0^\pi w_m(\phi) \,
(\vec r\cdot \vec n_\phi)^m \, d\phi,
\end{equation}
within the unit circle $\vert \vec r \vert \le 1$ using $K\ge M$
evenly spaced beams. Here $w_m(\phi)$ are arbitrary weight factors. Equation (\ref{EQ_APP2D}) can be considered as a
special form of a polynomial approximation of an arbitrary
two-dimensional dose distribution $D(\vec r)$.  A perhaps more
important result of that paper is that the required number of beams equals $M$, the
maximum degree polynomial plus 1. In particular, {\em there is no benefit at
all in using more than $M$ beams}. For a better understanding of
these finding I will consider a specific set of polynomials,
Chebyshev polynomials.

\subsection{Chebyshev polynomials of the first kind}
\begin{figure}[htbp]
\begin{center}
\includegraphics[width=10cm]{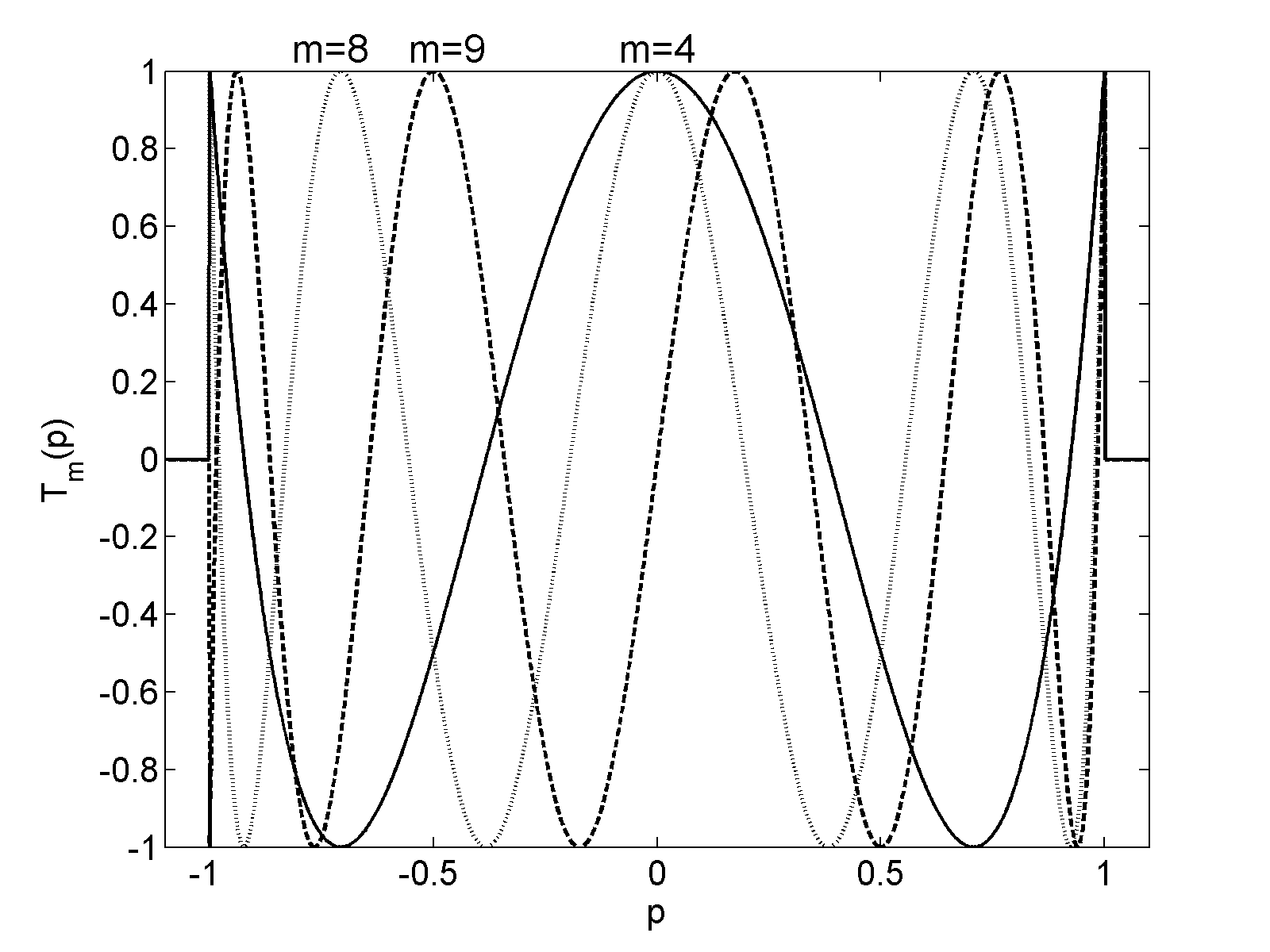}
\end{center}
\caption[tm]{\label{FIG_TM}
Examples of Chebyshev polynomials of the first kind, $T_m(p)$, for $m=4$ (solid line),
$m=8$ (dotted line), and $m=9$ (dashed line).}
\end{figure}

Chebyshev polynomials of the first kind, $T_m(p)$ are a special set
of (orthogonal) polynomials that can also be defined using cosine functions, namely:
\begin{equation}
T_m(p) = \cos(m \arccos p)
\end{equation}
They are defined on the interval $p \in [-1,1]$ and can be
considered as cosine or sine functions for even and odd $m$, respectively, with a frequency
that increases away
from the origin. On the other hand, they are also polynomials, for
example $T_4(p) = 8 p^4 - 8 p^2 + 1$,
$T_8(p) = 128 p^8 - 256 p^6 + 160 p^4 - 32 p^2 + 1$ and
$T_9(p) = 256 p^9-576 p^7 + 432 p^5 - 120 p^3 + 9p$. Figure \ref{FIG_TM} shows those
Chebyshev polynomials $T_4(p)$, $T_{8}(p)$, and $T_9(p)$.

In an earlier paper \cite{Bortfeld1999} some
properties of Chebyshev polynomials of the {\em second} kind,
$U_m(p)$, were utilized. Those Chebyshev polynomials of the second kind are
Eigenfunctions of the operation consisting of a backprojection and
subsequent projection under a different angle. The two kinds of Chebyshev polynomials
are related through the following equation:
\begin{equation} \label{EQ_U_T}
T_m(p) = \frac{U_m(p) - U_{m-2}(p)}{2}.
\end{equation}
Here I will focus on Chebyshev polynomials of the {\em first} kind
because they oscillate between $-1$ and $+1$, and issues with
negativity \cite{Cormack1989} can therefore be avoided simply by a zero-offset of $+1$.
Chebyshev polynomials have been used in the theory of IMRT by
Cormack et al. \cite{Cormack1987,Cormack1987a,Cormack1989} but, to the best of my knowledge,
they have not been used to get a handle on the beam angle problem.

\section{`Making' rotating beams from fixed beam angles}
The fact that one does not need more than $M$ beams if each of the
beam fluence profiles can be described by a polynomial of degree up
to $m=M-1$, will now be illustrated using the polynomial $T_4(p)$
(the solid line in figure \ref{FIG_TM}) as an
example. If one uses $K = 5$ fixed beams evenly spaced at angles
$\phi_k = 180^\circ k /K$, where $k=0,\ldots,K-1$, and each of those 5 beams delivers a $T_4$
fluence profile of variable weight, then one can `make' a beam with a
$T_4$ fluence profile from an {\em arbitrary} angle $\phi$. The weights of
the fixed beams to be used are given by
\begin{equation} \label{EQ_WM}
w(\phi_k) = \frac{1}{K} \,
\frac{\sin\left((m+1)(\phi_k-\phi)\right)}{\sin (\phi_k-\phi)},
\end{equation}
with $m=4$ and $K=5$ in this case. This result follows from appendix A in
\cite{Bortfeld1999}, in which $U_m$ has to be replaced with $U_{m-2}$, and then from
equation (\ref{EQ_U_T}). If $\sin (\phi_k-\phi) = 0$, then
$w(\phi_k) = \pm (m+1)/K$, and the sign can be determined with l'H\^opital's rule.

\begin{figure}[htbp]
\subfloat[]{\includegraphics[width=9cm]{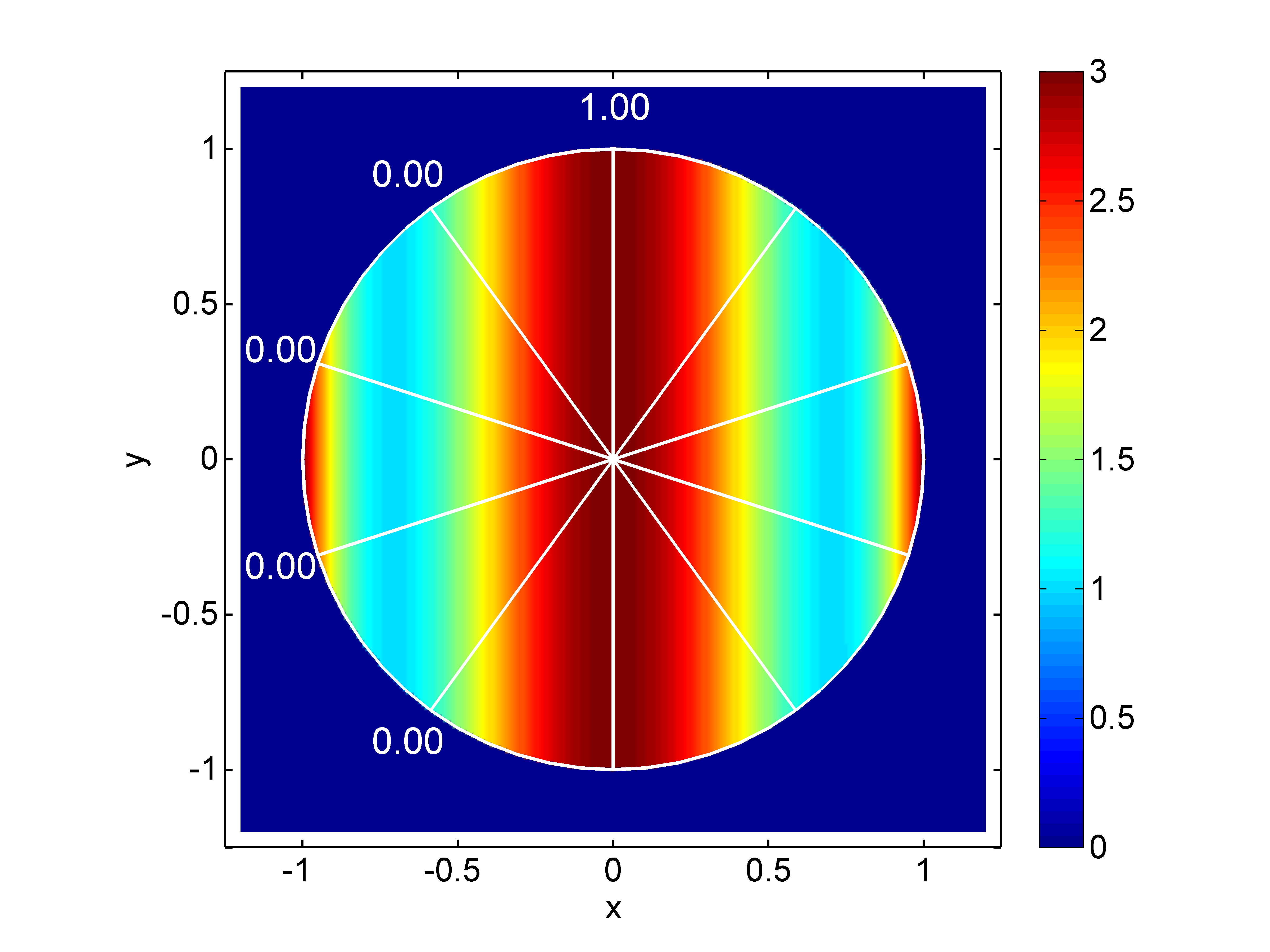}}
\subfloat[]{\includegraphics[width=9cm]{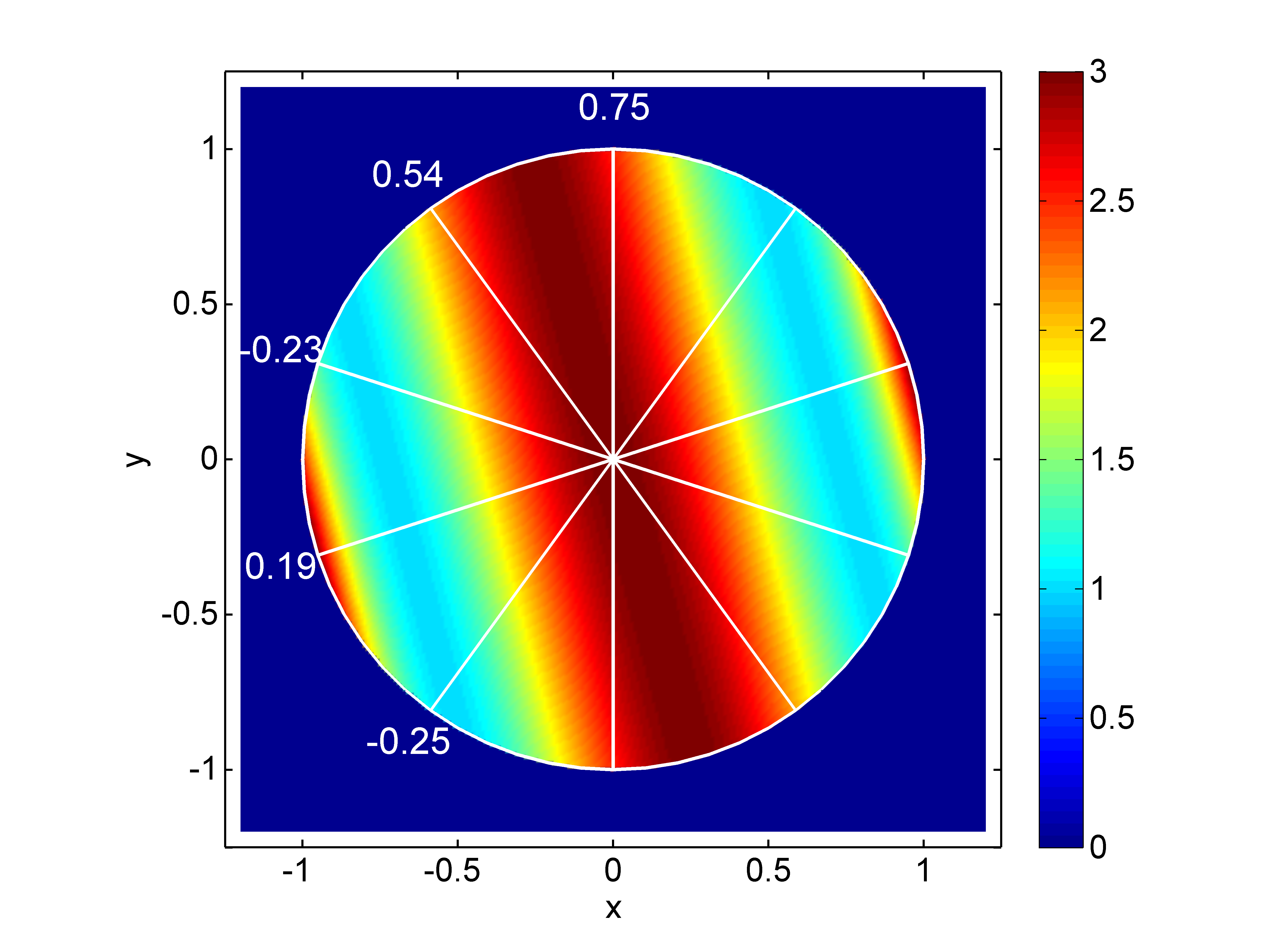}} \\
\subfloat[]{\includegraphics[width=9cm]{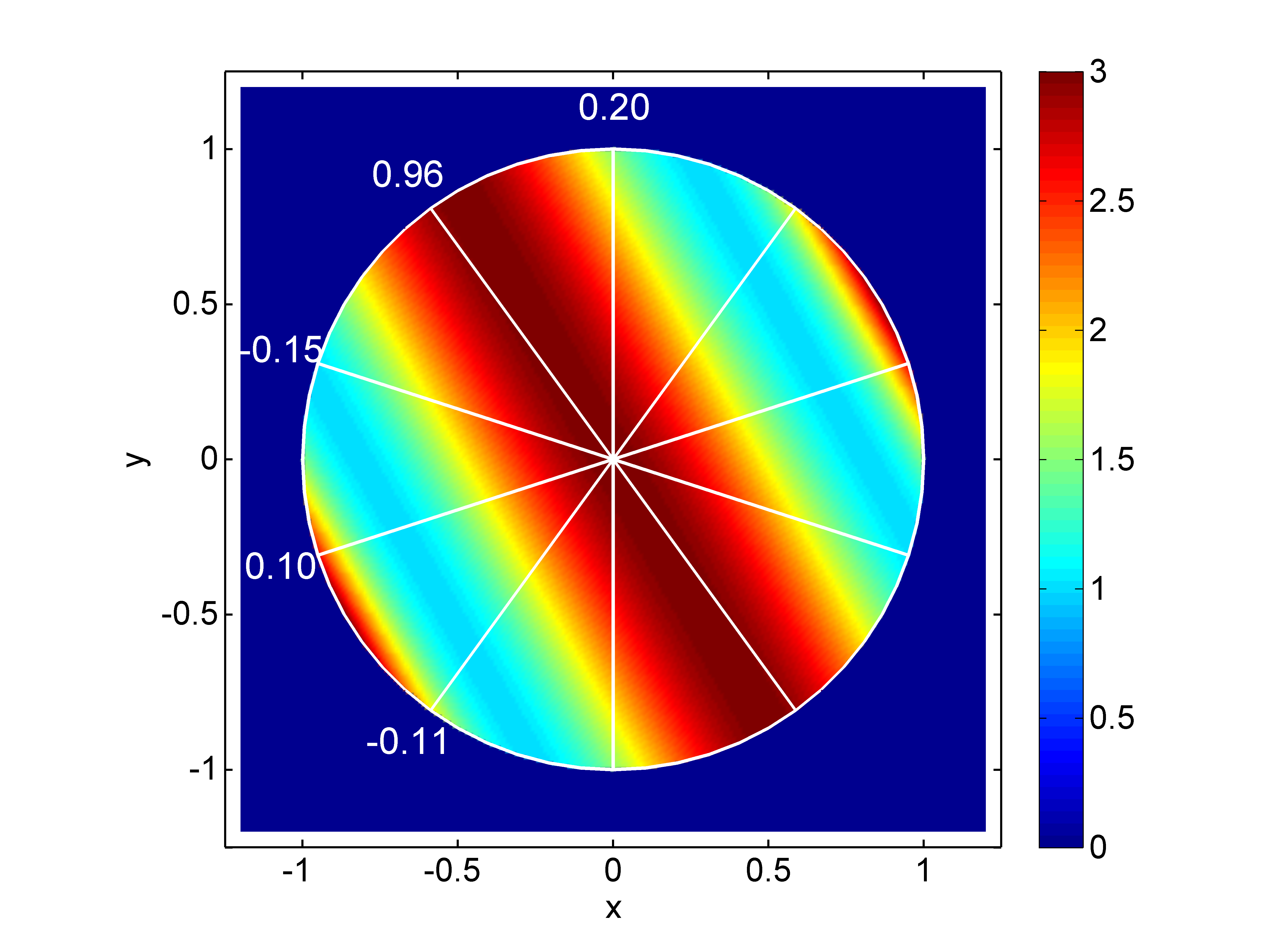}}
\caption[tm]{\label{FIG_TM_DOSE} Making rotating beams from fixed beam angles.
By varying the weights of intensity-modulated beams from fixed beam directions,
one can create beams from arbitrary directions, (a): $0^\circ$, (b): $15^\circ$,
(c): $30^\circ$. In every case, 5 beams from fixed evenly spaced directions (every $36^\circ$)
are used, as shown by the white lines. Every beam produces a Chebyshev $T_4(p)$ fluence
profile but with different weights that are calculated with equation (\ref{EQ_WM}) and
shown as white numbers next to the beams. Negative values of the fluence profiles are
avoided by a zero-offset, i.e. by adding constant fluence. See the text for further information.}
\end{figure}

Figure \ref{FIG_TM_DOSE} illustrates these findings graphically for
$\phi = 0^\circ$, $\phi = 15^\circ$, and $\phi = 30^\circ$. The
weights calculated with equation (\ref{EQ_WM}) are attached to the 5
beams in the plots. In figure \ref{FIG_TM_DOSE}a with the beam in vertical
direction ($\phi = 0^\circ$), only the vertical fixed beam ($\phi_0 = 0^\circ$)
has a weight of 1, and the four others a weight of 0. In figures \ref{FIG_TM_DOSE}b
and \ref{FIG_TM_DOSE}c all five beams contribute with a finite weight to `make'
those oblique beam directions.

For each of the beams, a zero-offset (i.e., a uniform fluence profile) has to be added to
make the fluence
non-negative. The total zero-offset over all beams depends on the weights $w(\phi_k)$,
as calculated using equation (\ref{EQ_WM}), and therefore on $\phi$. The biggest
total zero-offset is required for $\phi = 180^\circ / 2K = 18^\circ$, and is close to 2.
I therefore use a total zero-offset of 2 in all cases shown in figure \ref{FIG_TM_DOSE}.
I should mention that a $T_4$ dose profile with arbitrary orientation
can also be generated from 5 fixed beams with intensity profiles that are polynomials
of degree 4 but not Chebyshev $T_4$ profiles. In other words, there is a `nullspace'
that leads to non-uniqueness \cite{Cormack1987}. However, one can show numerically that
the use of 5  $T_4$ profiles (evenly spaced) requires the smallest zero-offset.

Moreover, it is interesting to note that beams with arbitrary orientation can also be
generated from a number of fixed beams that are {\em not} evenly spaced but arbitrarily located.
This finding may serve as an explanation that the choice of beam angles is generally less
critical in IMRT than in standard 3D conformal radiation therapy without intensity modulation.
However, with non-equally spaced beams, the solutions (the equivalent of equation (\ref{EQ_WM}))
may become highly oscillatory (i.e., one beam may require a large positive weight and the
next a large negative weight), and very large zero-offsets may be required.

\section{A simple two-dimensional example}
\label{SEC_2DEXAMPLE}
For a further illustration of the findings from above, let us now
look at a simple 2-D IMRT example case: a circular target volume that `wraps around' a
circular critical structure. To match the region of support of the Chebyshev polynomials,
the radius of the target is set to 1 unit, with its center positioned at $(x_0,y_0) = (0,0)$.
The critical structure has a radius of 1/3 units and its center is at $(0,-2/3)$. This
geometry corresponds with the geometry used in a recent paper \cite{Bortfeld2009}. The
dose prescription has a dose level of 1 unit to the target volume, and a reduction to 0.5
units in the critical structure. The 50\% dose reduction in the critical structure is
chosen as a challenging but realistic value, and to avoid any issues with negativity.

The calculation of the IMRT fluence profiles is done with the recipe from an earlier paper,
see page 1109 in \cite{Bortfeld1999}. Here I approximate the fluence profiles
with $M=10$ Chebyshev polynomials with degrees between $m=0$ and $m=9$, see figure \ref{FIG_TM}.
The specific steps of the calculation are:
\begin{enumerate}
\item Projection of the prescribed dose along $K$ evenly spaced beam directions
with angles $\phi_k$ at lateral
positions $p_j = \cos\!\left(\frac{\pi j}{M+1}\right), \, \, j=1,\ldots,M$.

\item Discrete sine transform of each of those $K$ projections.

\item Multiplication of the coefficients with a `ramp filter'.

\item Composition of `filtered projections' to obtain the $K$ fluence profiles.

\end{enumerate}
The last step above involves the use of Chebyshev polynomials of the {\em second} kind.
For consistency with the rest of this paper, the algorithm can be re-written
for the use of Chebyshev polynomials of the first kind through equation (\ref{EQ_U_T}).

\begin{figure}[htbp]
\begin{center}
\subfloat[]{\includegraphics[width=10cm]{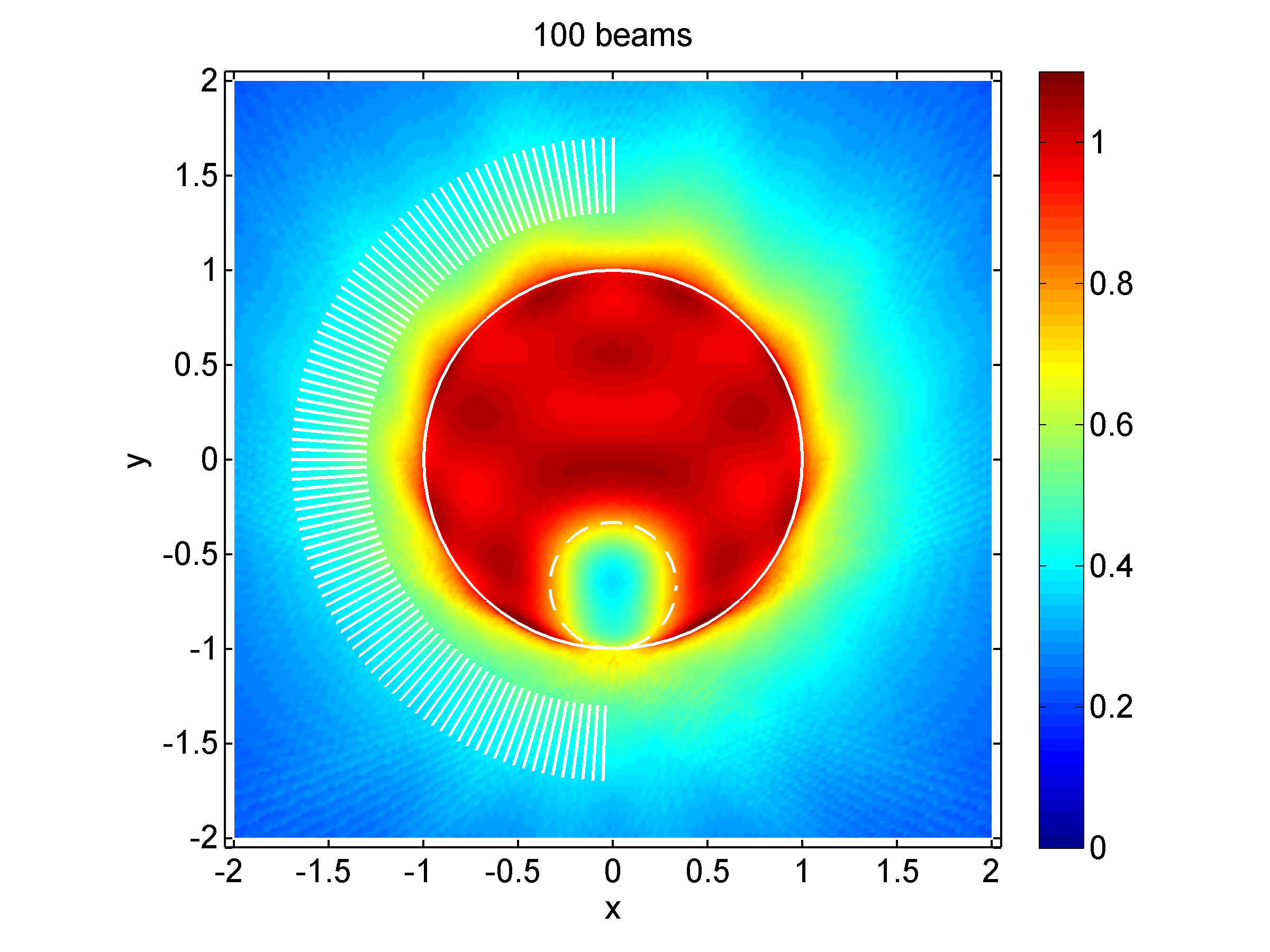}} \\
\subfloat[]{\includegraphics[width=10cm]{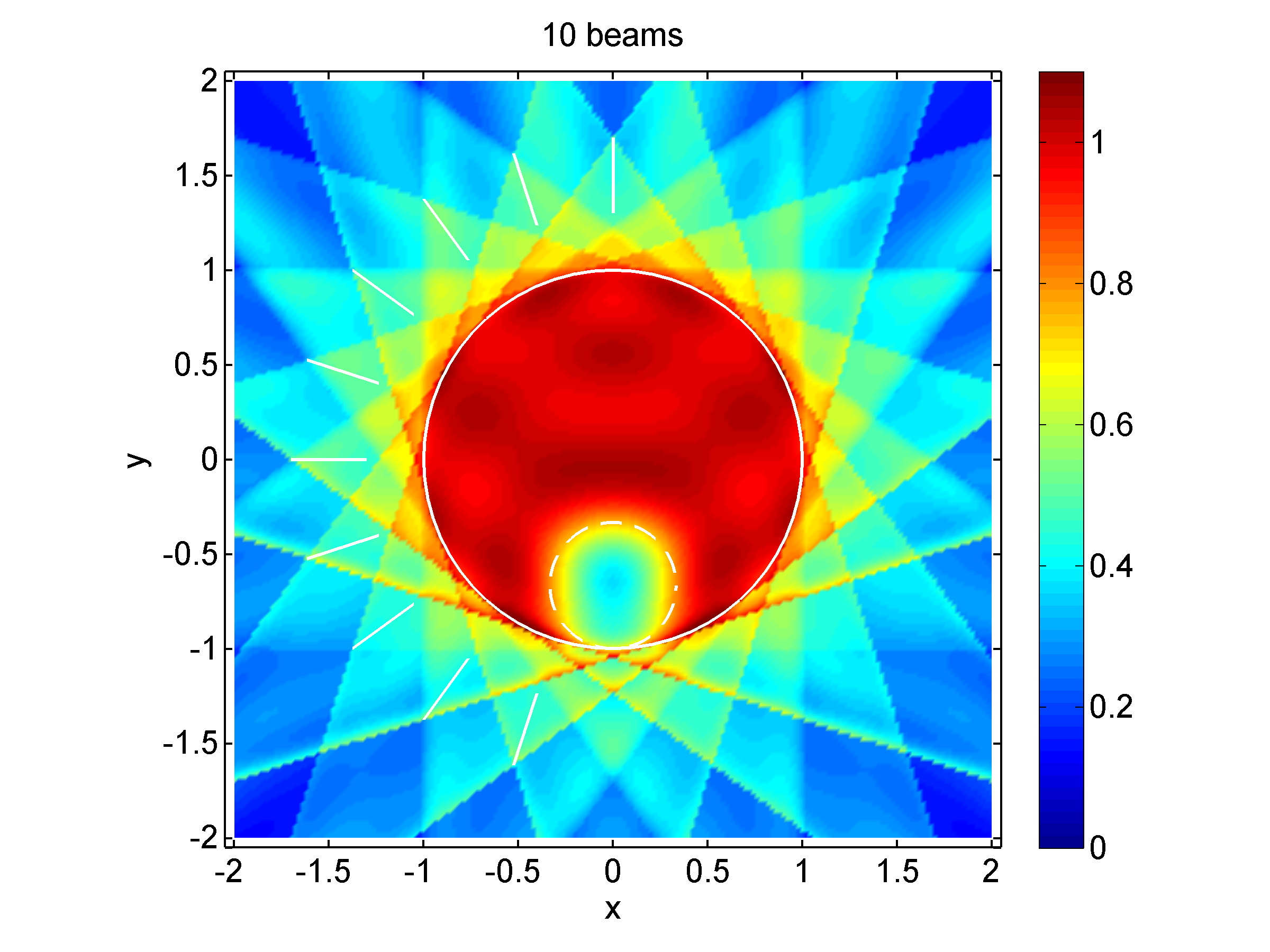}}
\end{center}
\caption{\label{FIG_CIRC} This figure shows dose distributions for the treatment
of a circular target volume (white solid line) that `wraps around' a small
circular critical structure (white dashed line) with $K=100$ beams (a) and $K=10$ beams (b).
The dose prescription is one dose unit to the target volume, and a dose reduction to 0.5
units is requested in the critical structure. All beams are composed of $M=10$ Chebyshev
polynomials with degrees between $m=0$ and $m=9$. Within the target circle, which
encompasses the critical structure, the two distributions in (a) and (b) are {\em identical},
i.e., there is no benefit in using more than 10 beams here.}
\end{figure}

After the calculation of the $K$ fluence profiles, those profiles are then backprojected,
which simulates an irradiation with $K$ beams. Figure \ref{FIG_CIRC}(a) shows the resulting
dose distribution for $K=100$ beams that are evenly spaced at every $1.8^\circ$. Now, as
I said above, for polynomial fluence profiles of degree up to $m=9$, $K=10$ beam angles
should actually suffice. To illustrate this point, I use $K=10$ beams at every $18^\circ$
in figure \ref{FIG_CIRC}b. The 10 fluence profiles are calculated from the case with 100
beams by use of equation (\ref{EQ_WM}). I should note that after the re-distribution into
the 10 fluence profiles, there is no guarantee that all of them are non-negative, even if
the original 100 profiles are all non-negative. However, in this specific case negativity
is not an issue.
By comparing figures \ref{FIG_CIRC}(a) and \ref{FIG_CIRC}(b) one can now
see that the two dose distribution are indeed {\em identical} within the unit circle,
which encompasses the target volume and the critical structure. Only outside of the unit
circle the two distributions begin to differ more and more: with $K=100$ beams the distribution
is more uniform, whereas the  case with $K=10$ beams exhibits the typical ragged
distribution. With fewer than $M=10$ beams, there would be differences also within
the unit circle.

\section{How many beams are needed?} \label{SEC_HOWMANY}
Can one now find a general answer to this question?
As I said above, the variability of the fluence profiles, measured by the degree of the
Chebyshev polynomials approximating those fluence profiles, determines how many beam are needed.
The fluence variability depends, in turn, on the complexity of the dose prescription,
and is therefore case dependent. However, even the most complex cases do not benefit
from an arbitrarily high number of beams. This is so because there is also a physical limit
on the achievable amount of  fluence (or rather dose) modulation, which is due to the reasons
mentioned above (scatter, transmission, finite source size). This physical limit reveals itself
through the fact that the dose fall-off at a field edge is never infinitely steep, but
exhibits a dose `penumbra'. Specifically, the width of the penumbra (80\% to 20\%) near
the leaf-end of an MLC at a typical treatment depth is of the order of 7 mm \cite{Boyer1992}.
As a consequence, the minimum width (full width at half maximum, FWHM) of a deliverable dose
peak, $\Delta_p$, is of the order of $1.4 \times 7\mbox{ mm} = 1 \mbox{ cm}$
\cite{Bortfeld2000}. Now, by interpreting Chebyshev fluence profiles  as a series of dose
peaks (see figure \ref{FIG_TM}), one can correlate this physical limit with the maximum degree
polynomial that can be delivered, and thus with the required number of beams.

Chebyshev polynomials $T_m(p)$ of the first kind have $m$ roots. The roots
cut the $[-1,1]$ interval into $m+1$ sub-intervals, which represent the peaks and
valleys of $T_m(p)$. The average width of the sub-intervals is $2/(m+1)$, hence the
average peak width is also $2/(m+1)$ (FWHM). To give this expression a quantitative
meaning, one has to identify the 2 units in the numerator with $2R$, where $R$ is the
radius of the the circular region of interest, which was set to 1 in the abstract
examples above. The region of interest contains the target volume and the most important
nearby critical structures, which may extend into the target volume. Thus the average
peak width is $2R/(m+1)$. If one equates this with the physical limit, $\Delta_p$ above,
and utilizes the fact that $m+1$ governs the required number of beams, $K$, one obtains:
\begin{equation} \label{EQ_KMIN}
K \approx \frac{2R}{\Delta_p}.
\end{equation}
To give a couple of examples, for relatively small cases where the radius of the region of
interest is $R=5\mbox{ cm}$, $K=10$ beams are required, whereas for larger cases with
$R=10\mbox{ cm}$,
one needs $K=20$ beams. For `sharper' beams with smaller penumbra, i.e., with a
smaller $\Delta_p$, the number is correspondingly higher.

I should emphasize that equation (\ref{EQ_KMIN}) can only serve as a rough estimate
of the required number of beams, as will be discussed in section \ref{SEC_DIS}.

\section{Towards a theoretical justification of Single-Arc}
Now that it has been shown that a finite and moderate number, $K$, of
fixed IMRT beams suffices, how can one justify Single-Arc IMRT (RapidArc, VMAT),
in which the number of beams is effectively infinite, but without intensity modulation
from any one beam direction? One way to provide a theoretical foundation of Single-Arc
is as follows: Let us start with $K$ beams and assume that the fluence profile for each
beam is delivered with a `sweep' technique \cite{Convery1992}, see \ref{SEC_APP},
in which the leaves move uni-directionally, say from left to right. What is the
impact on the fluence and dose distribution, if one delivers the dose from each beam
not while the gantry is static at a fixed beam angle, but rotating over the angle
$180^\circ / K$? Some investigations to answer this question were presented in \cite{Webb2009}.
 A closed form answer is derived in \ref{SEC_APP}. Basically the
impact of gantry motion during leaf sweep IMRT delivery is a stretching or compression
of the fluence profile\footnote{This effect
is quite different in the case of beam modulator (`compensator') based IMRT and tomotherapy,
in which gantry motion leads to a `smearing' of the fluence distribution.}, depending on
whether the gantry motion increases (prograde motion) or
decreases (retrograde motion) the motion of the field edges (the projected leaf ends).

For leaf sweep IMRT, the fluence profiles $f(p)$ are typically decomposed into two
components $f^+(p)$ and $f^-(p)$, which represent the positive and negative slopes
in $f$, respectively, and for which $f(p) = f^+(p) - f^-(p)$. Now let
$\tilde f^+(p)$ and $\tilde f^-(p)$ be the respective fluence profiles that
result when there is gantry motion during delivery. In the case of prograde motion,
where the effective leaf speed is increased by a positive value $v$, it is shown in \ref{SEC_APP1}
that for any point $p = p_0$
\begin{equation} \label{EQ_PROG_TEXT}
\tilde f^\pm\left(p_0 - \Delta p + v f^\pm(p_0) \right) =  f^\pm(p_0),
\end{equation}
where $\Delta p$ is an arbitrary constant shift, which could be zero. This means
that the $\tilde f^\pm$ profile has the same values as the $f^\pm$ profile, but it
is stretched in the direction of $p$.
For retrograde motion, where $v$ is negative, the corresponding equation (\ref{EQ_RET})
is derived in \ref{SEC_APP2} and looks a little more complicated because leaves can move
backwards in this case.
However, because the stretching or compression is generally different
for $\tilde f^+$ and $\tilde f^-$, the impact on $\tilde f = \tilde f^+ - \tilde
f^-$ can be more substantial.

\begin{figure}[htbp]
\begin{center}
\subfloat[]{\includegraphics[width=10cm]{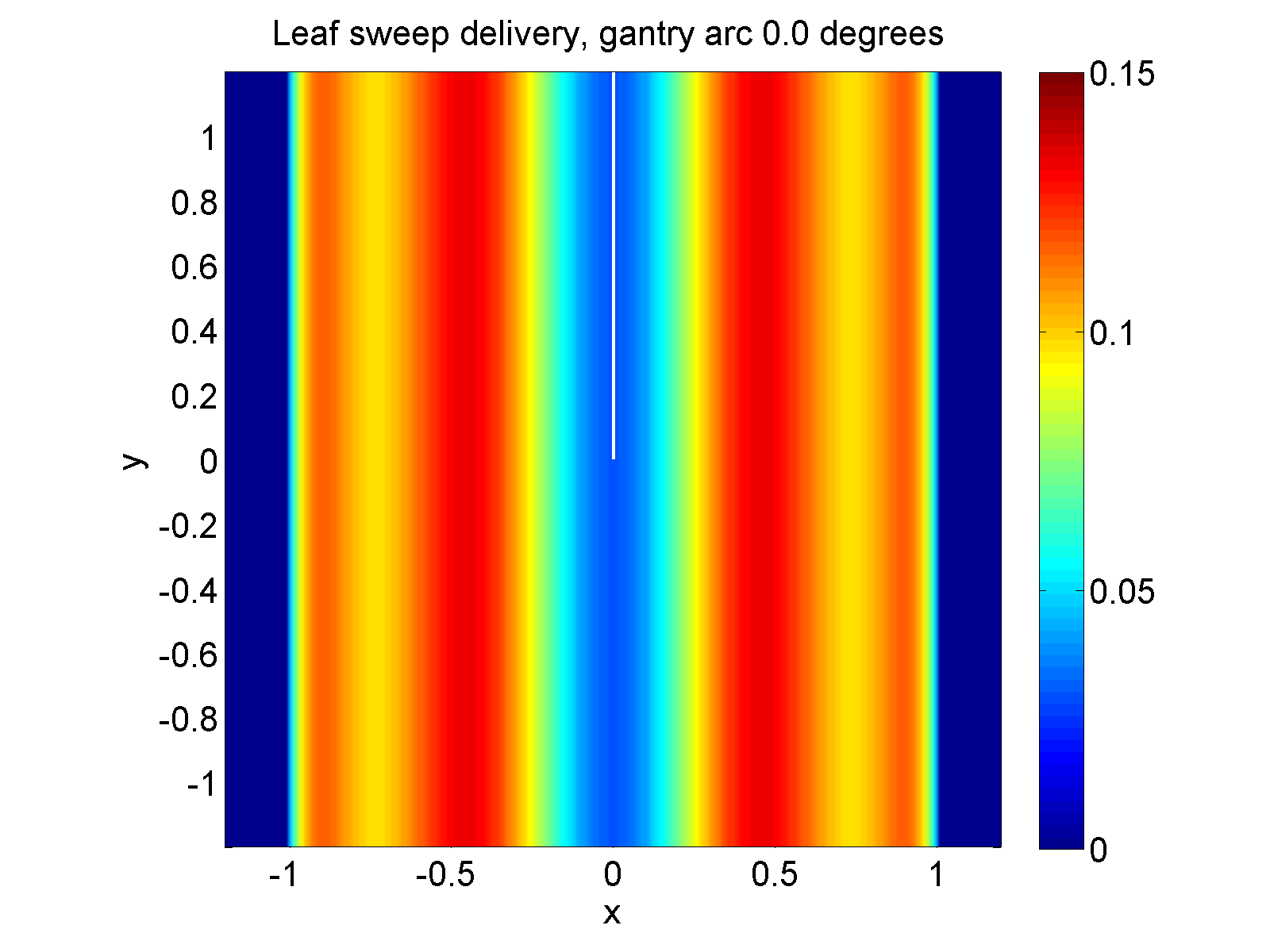}} \\
\subfloat[]{\includegraphics[width=10cm]{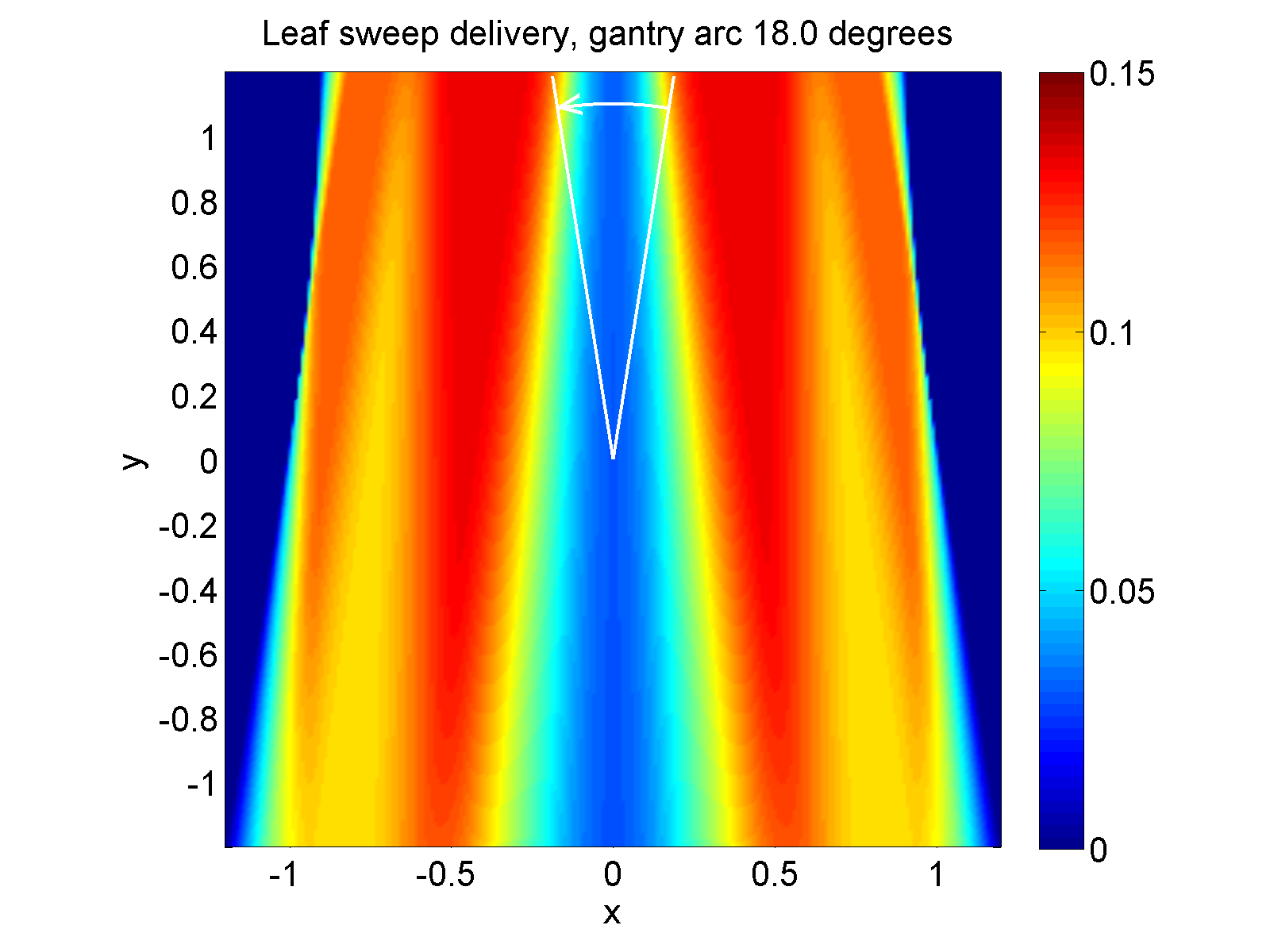}}
\end{center}
\caption{\label{FIG_VMAT} The effect of gantry motion during leaf sweep IMRT delivery
is illustrated here for one of the 10 beams that produce the dose distribution shown in
figure \ref{FIG_CIRC}(b). Specifically, the vertical beam ($\phi_0=0^\circ$)
from that example is considered. The dose from that beam without motion during delivery is
shown in (a)
above. Here the beam comes straight from the top. In (b) gantry motion is simulated in
counter-clockwise direction from $\phi = -9^\circ$ to $9^\circ$ during leaf sweep IMRT
delivery of that beam. The leaves move from the left to the right. In the upper
half of part (b) (for positive $y$), the resulting motion of the field edges in the $x$-direction
is {\em reduced} by the gantry motion (retrograde motion),
whereas in the lower half (negative $y$) it is increased (prograde motion). At $y=0$
there is no effect of the gantry motion.}
\end{figure}

Let us now illustrate the findings above for the example case shown in figure \ref{FIG_CIRC}.
The dose distribution $D_0(x,y)$ from the first of the 10 beams used in \ref{FIG_CIRC}(b)
(the vertical one at $\phi_0=0^\circ$), is shown in figure \ref{FIG_VMAT}(a). It is composed
of the following Chebyshev polynomials:
$D_0(x,y) = (0.935 T_0(x) - 0.018 T_2(x) -0.312 T_4(x) + 0.076 T_6(x) - 0.262 T_8(x))/10$.
Note that there are only polynomials of even degree in this beam profile because of the
left-right symmetry of the example case.
The effect of gantry motion is simulated over an arc segment of $18^\circ$,
from $-9^\circ$ to $9^\circ$, during the delivery of this beam. The gantry
rotates counter-clockwise. The leaf motion is from left to right. Thus, in the upper
half of figure \ref{FIG_VMAT}(b), the resulting motion of the field edges in the $x$-direction
is {\em reduced} by the gantry motion (retrograde motion),
and the effect is modeled with equation (\ref{EQ_RET}) with negative $v$.
In the lower half, which exhibits prograde motion, equation  (\ref{EQ_PROG_TEXT}) with positive $v$
is applied.

\begin{figure}[htbp]
\begin{center}
\includegraphics[width=10cm]{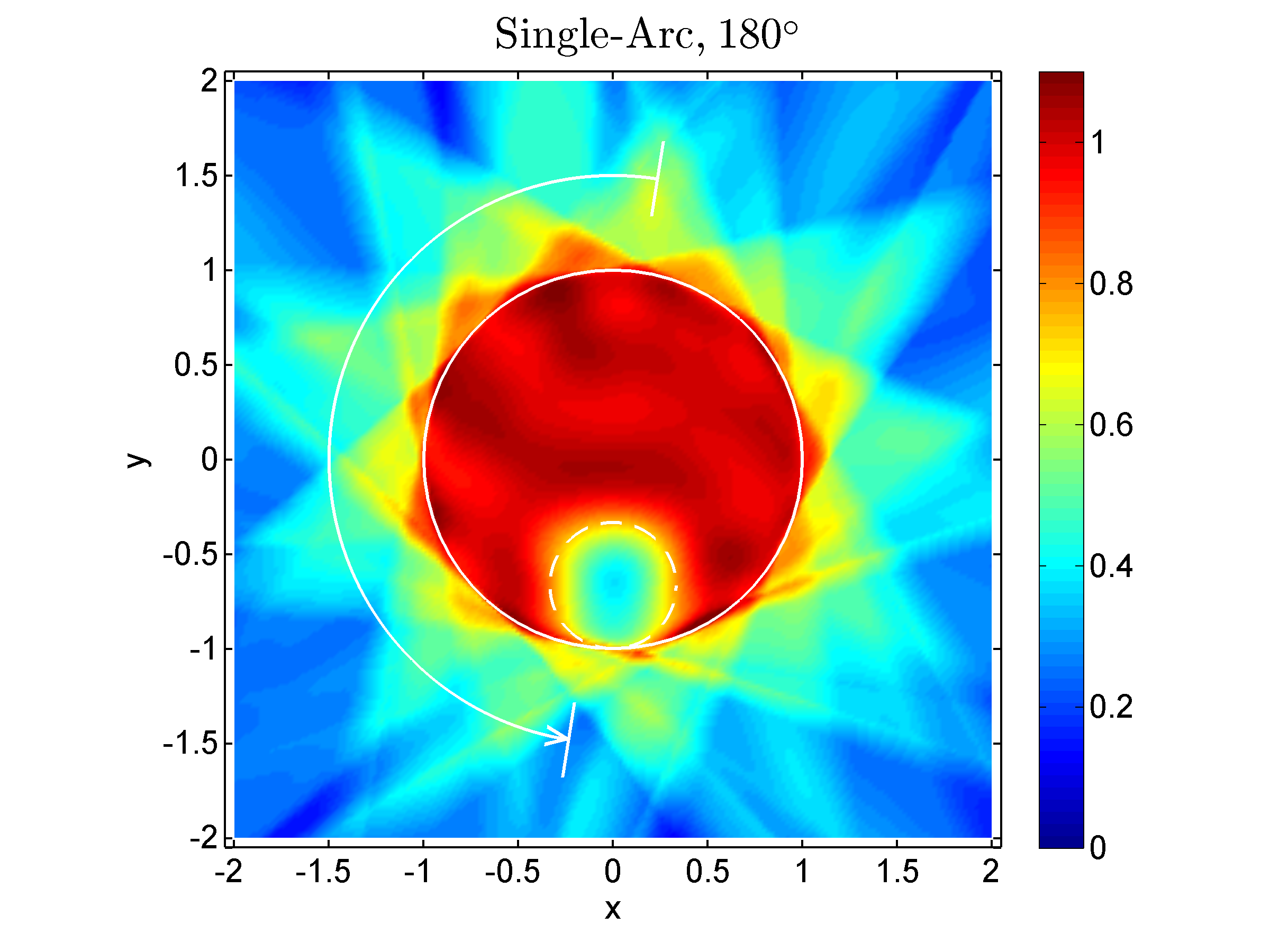}
\end{center}
\caption{\label{FIG_CIRC_VMAT} This figure shows the dose distribution that results
from delivering the 10 beams used in figure \ref{FIG_CIRC}(b) not from static gantry
angles but during continuous gantry motion in the Single-Arc mode. The first of the 10
beams is delivered with MLC leaf sweep while the gantry rotates from $-9^\circ$ to
$9^\circ$, the second during the gantry arc segment from $9^\circ$ to $27^\circ$, and
so forth. The total Single-Arc is being delivered over $180^\circ$, from $-9^\circ$
to $171^\circ$.}
\end{figure}

The overall result is that, with motion, the characteristics of the fluence profile are preserved,
and the effect of motion appears to be similar to using a divergent beam.
In Single-Arc, the effect is reduced because the MLC moves back and forth during a rotation of
the gantry, i.e., there is an alternation between retrograde and prograde motion. The total
dose distribution, which results from combining the 10 arc segments according to the 10 beams
used in figure \ref{FIG_CIRC}(b), is shown in figure \ref{FIG_CIRC_VMAT}. Comparing the two
figures (\ref{FIG_CIRC}(b) and \ref{FIG_CIRC_VMAT}), one can see that the effects on the dose
distribution within the unit circle encompassing the target and the critical structure are small.
A numerical analysis reveals that the average (RMS) difference is 0.02, i.e., 2\% of the
prescribed dose. Outside of the unit circle, some noticeable smoothing occurs in Single-Arc,
and the differences are bigger. One can also see that the Single-Arc solution is slightly
asymmetric. Incidentally, if an odd number of beams had been used in this example, the solution
would have been perfectly left-right symmetric.

\section{Discussion} \label{SEC_DIS}
In this paper an attempt was made to approach the beam angle problem
in IMRT from a fundamental theoretical direction. The objective was to
derive analytical expressions that would allow one to draw some
generally applicable, rather than case-specific, conclusions. Indeed
this objective has been met. However, it could only be accomplished at
the price of a strongly simplified model. The question arises: what is
the validity of the findings in the real world? I claim that the
results remain valid if one uses more realistic dose
models, and provide the following arguments in support of this
statement. The physical effects that were neglected in the
oversimplified model, i.e., dose build-up, (exponential) dose fall-off, and beam divergence, lead to {\em smooth} variations from the model.
The real world can therefore be modeled by adding small-degree
polynomials to the simple model. As long as the degree of these
correction polynomials does not exceed the maximum degree polynomial
used in the model, the conclusions remain valid.

With respect to the determination of the required number of beams,
I should emphasize that equation (\ref{EQ_KMIN}) can only serve as a rough estimate. First, in the interpretation of the Chebyshev
polynomials as a series of peaks, I ignored the variation of the peak widths
(see figure \ref{FIG_TM}) and just considered the average width. A more careful
analysis would have to take the variation in width into account. Secondly, strictly
speaking equation (\ref{EQ_KMIN}) is only valid if one allows negative beam intensities.
I address the negativity issue by use of the zero-offset idea mentioned above.
However, the total zero-offset depends on the number of beams.
A more detailed analysis would have to take this second-order effect,
which was also mentioned in section \ref{SEC_2DEXAMPLE} above, into account.
Third, the number calculated with equation (\ref{EQ_KMIN}) is not a {\em strict}
practical limit on the number of beams. In fact, smaller numbers of beams suffice
to obtain approximate solutions. Along similar lines, it is clear that the required number of beams may also depend on the specific geometry of the target volume and the surrounding critical structures, i.e, simple cases require less beams. This paper does not provide a case-specific analysis of the number of beams, because a general answer is sought after.

As far as Single-Arc is concerned, a potential criticism of this work is that the theory is only two-dimensional. This would not be a problem as long as the real 3D
solution can be composed of 2D slices. However, in a 2D model one
cannot model collimator rotation. Others have found that in Single-Arc a 45$^\circ$
collimator rotation leads to better results \cite{Otto2008} than if
the MLC leaves move within the transversal planes, as in this paper. It
is not completely clear where the advantage of a 45$^\circ$ collimator comes
from. It has been hypothesized that this advantage is due to the fact
that, with the collimator at 45$^\circ$, in parallel opposed beams the leaves of the MLC move
in orthogonal directions. In other words, parallel opposed beams
are not redundant (unlike in the present model with a flat dose profile, where they
would indeed be redundant). It
should be kept in mind, however, that the model here assumes a gantry rotation
of only about 180$^\circ$. Hence, the delivery could be faster, maybe
twice as fast, as a Single-Arc treatment over 360$^\circ$ with a 45$^
\circ$ collimator rotation. The 45$^\circ$ collimator rotation may be
of no advantage if the Single-Arc angle is restricted to 180$^\circ$.

With respect to future developments, the idea to decompose beam profiles into well-behaved smooth functions, such as Chebyshev polynomials, may not only be of interest for the analysis of the required number of beams, but it could also be beneficial for general IMRT treatment planning and optimization.

\section{Conclusions}
In IMRT the required number of beams depends directly on the complexity of the fluence
(intensity) profiles that can be delivered within the physical and technical constraints of the treatment machine. One can measure the fluence complexity by the degree
of the Chebyshev polynomial that is required to approximate the profile.
The required number of beams then equals the highest degree polynomial plus 1.
Rotating the beams over short arc segments between the fixed beam positions merely
causes the dose distribution to fan out, but does not normally alter it substantially.
This, in combination with the first conclusion above,  serves as a theoretical justification
of single-arc IMRT (VMAT).

\begin{appendix}

\section{The impact of gantry rotation on leaf sweep IMRT delivery}
\label{SEC_APP}
Let us first review the mathematical basis of  leaf sweep MLC
delivery.
Consider a (non-negative) fluence profile $f({p})$ in the interval $ [{p}_a,{p}_b]$
such that $f({p}_a) = f({p}_b) = 0$. One can always decompose $f$ into two
components, one of which, $f^+$, contains all positive slopes and the other,
$f^-$, contains all negative slopes:

\begin{equation}
f({p}) = f^+({p}) - f^-({p})
\end{equation}
where
\begin{equation}
f^+({p}) = \int\limits_{{p}_a}^{{p}} \left[\frac{df}{d{p}} \right]^+ d{p} \end{equation}
and \begin{equation}
f^-({p}) = \int\limits_{{p}_a}^{{p}} - \left[\frac{df}{d{p}} \right]^- d{p}.
\end{equation}

Here $[g]^+$ equals $g$ if $g>0$, and 0 otherwise. Similarly, $[g]^-$ equals $g$
if $g<0$, and 0 otherwise. Note that both $f^+({p})$ and $f^-({p})$ are
monotonically increasing functions.

Let us first consider $f^+({p})$. For any argument ${p}={p}_0$ one can write $f^+({p}_0)$
in the form:
\begin{equation}
f^+({p}_0) = \int\limits_{{p}_a}^{{p}_b} f^+({p}) \delta({p}_0 - {p}) d{p} = \int\limits_{{p}_a}^{{p}_b} f^+({p}) \delta({p} - {p}_0) d{p}.
\end{equation}
Using integration by parts, i.e., $\int F g \, d{p} = FG - \int fG \, d{p} $, one  obtains further:
\begin{equation}
f^+({p}_0) = \left[f^+({p}) H({p} - {p}_0) \right]_{{p}_a}^{{p}_b>{p}_0}
- \int\limits_{{p}_a}^{{p}_b} \frac{df^+}{d{p}} H({p} - {p}_0) d{p},
\end{equation}
because the integral of a delta function is the Heaviside step function, $H$. With the identity $H({p}-{p}_0) = 1 - H({p}_0-{p})$ one can then easily show that
\begin{eqnarray}
f^+({p}_0) &= \int\limits_{{p}_a}^{{p}_b} \frac{df^+}{d{p}} H({p}_0 - {p}) d{p} \\
&= \int\limits_0^{f_{\max}} H\left({p}_0 - {p}^+(f)\right) df. \label{EQ_fplus_0}
\end{eqnarray}

The case for $f^-({p}_0)$ can be made completely analogously (simply replace $f^+$
with $f^-$ and ${p}^+$ with ${p}^-$), yielding:
\begin{equation}
f({p}_0) = f^+({p}_0) - f^-({p}_0) = \int\limits_0^{f_{\max}} H\left({p}_0 -
{p}^+(f)\right) - H\left({p}_0 - {p}^-(f)\right) df.
\end{equation}
This is indeed the standard uni-directional leaf sequencing algorithm in which
the trajectory of the left (A) leaf is given by ${p}^+(f)$ and the trajectory of
the right (B) leaf by ${p}^-(f)$. Here one should identify $f$ with time, $t$.

\subsection{Prograde motion}
\label{SEC_APP1}
For the sake of simplicity I will assume that gantry motion adds constant speed, $v$, to the speed of the leaves. Variable speed can be considered but at the expense of a more complex notation. Let us also consider a constant shift, $\Delta {p}$. In the following, the tilde stands for gantry motion, so:
\begin{equation} \label{EQ_PROGRADE}
\tilde {p}^\pm(f) = {p}^\pm(f) - \Delta {p} + v f.
\end{equation}
Consider first the case where the gantry motion is in the direction of the
moving MLC, that is, in the $+{p}$ direction, such that $v$ is positive.
From equation (\ref{EQ_fplus_0}) (tilde version) one obtains, using integration by parts:
\begin{eqnarray}
\tilde f^+({p}_0) &= \int\limits_0^{f_{\max}} H\left({p}_0 - \tilde {p}^+(f) \right) df \\
& =  \int\limits_{{p}_a}^{{p}_b} H\left({p}_0 - \tilde {p}^+({p}) \right) \frac{df^+}{d{p}} d{p} \\
& = \left[H({p}_0-\tilde {p}^+({p}))\, f^+({p}) \right]^{{p}_b}_{{p}_a} \nonumber \\
& \quad \quad - \int\limits_{{p}_a}^{{p}_b} \delta({p}_0-\tilde {p}^+({p}))
 \left(-1 - v \frac{df^+}{d{p}}\right)  f^+({p}) \, d{p}.  \label{EQ_tilde_f_1}
 \end{eqnarray}
Here $\tilde {p}^+({p})= {p} - \Delta {p} + v f^+({p})$ and
$ \left( -1 - v\frac{df^+}{d{p}} \right) = -\frac{d\tilde {p}^+}{d{p}}$. The first term
in equation (\ref{EQ_tilde_f_1}) is zero.
Now use the relationship $\delta (g({p})) = \sum_i \frac{\delta({p}-{p}_i)}{\vert d g/ d {p} ({p}_i) \vert} $, where ${p}_i$ are the roots of $g({p})$. In this case there is one root of ${p}_0 - \tilde {p}^+({p})$, which is
${p} = (\tilde {p}^+)^{-1}({p}_0)$. The denominator of the relationship for $\delta(g({p}))$ cancels out the term $ -\left( -1 - v\frac{df^+}{d{p}} \right)$ in equation (\ref{EQ_tilde_f_1}). This leaves us with:
\begin{eqnarray}
\tilde f^+({p}_0) &= \int\limits_{{p}_a}^{{p}_b} \delta\left({p} - (\tilde {p}^+)^{-1}({p}_0) \right) \, f^+({p}) \,  d{p} \\
&= f^+ \left( (\tilde {p}^+)^{-1}({p}_0) \right),
\end{eqnarray}
which can be written in the form:
\begin{equation}
\tilde f^+\left(\tilde {p}^+ ({p}_0)\right) =
\tilde f^+\left({p}_0 - \Delta {p} + v f^+({p}_0) \right) =  f^+({p}_0).
\end{equation}
Similarly,
\begin{equation}
\tilde f^-\left({p}_0 - \Delta {p} + v f^-({p}_0) \right) = f^-({p}_0).
\end{equation}This means that $\tilde f^\pm$ is just $f^\pm$ with variable
shifts in the direction of ${p}$. However, because those shifts can be different
for $\tilde f^+$ and $\tilde f^-$, the impact on $\tilde f = \tilde f^
+ - \tilde
f^-$ can be more substantial.

An even simpler  way to understand the impact of gantry motion is by looking at
the gradient of the fluence profile. From equation (\ref{EQ_PROGRADE}) one can see that
\begin{equation} \frac{d \tilde {p}^\pm}{df} = \frac{d {p}^\pm}{df} + v
\end{equation} or \begin{equation} \frac{d \tilde f^\pm}{d {p}} =  \left. \frac{d
f^\pm}{d {p}} \middle/
\left(1 + v
\frac{d f^\pm}{d{p}} \right) \right.
\end{equation}
Hence, in areas with large gradients of $f^\pm$, the gradient of $ \tilde f^\pm$
will converge to $1/v$. Areas with small gradients will not be affected.

\subsection{Retrograde motion}
\label{SEC_APP2}
Now let us look at the case where the gantry motion is in the opposite direction
of the moving MLC, that is, in the $-{p}$ direction, which means that $v$ is
negative. This is more complicated because now the MLC leaves can move in both
directions, i.e., $\tilde {p}^\pm (f)$ is no longer necessarily monotonic. As one
can see from equation (\ref{EQ_PROGRADE}), leaves move backwards, i.e.,  $\tilde
{p}^\pm (f)$ has a negative slope (and so does, consequently, $\tilde
{p}^\pm ({p})$) if $df^\pm/ d{p} > -1/v$. Let us first look at
$f^+({p})$ again.

From equation (\ref{EQ_tilde_f_1}) it follows that in this case
\begin{equation}
\tilde f^+({p}_0) = f_{\max} H\left({p}_0 - \tilde {p}^+({p}_b)\right) + \sum_i s_i f^+({p}_i),
\end{equation}
where the ${p}_i$ are, again, the roots of $g({p}) = {p}_0-\tilde {p}^+({p})$, and $s_i = -1$ if the slope of $\tilde {p}^+({p})$ is negative at ${p} = {p}_i$, and $s_i = +1$ otherwise.

Now the task is  to calculate $f^+({p}_i)$.
One can  keep this problem tractable by subdividing the fluence interval
$[0,f_{\max}]$ into $n$ contiguous sub-intervals $[\hat f^+_i,\hat f^+_{i
+1}]$ such that $\hat f^+_1 = 0$ and $\hat f^+_{n+1} = f_{\max}$. The
sub-intervals
are chosen such that in the $i$-th sub-interval $\tilde {p}^+ (f)$, and thus $\tilde {p}^+ ({p})$, is either
monotonically increasing or decreasing, in an alternating sequence.
The corresponding ${p}$ intervals are $[\hat {p}^+_i, \hat {p}^+_{i+1}], i = 1, \ldots, n$, where
$\hat {p}^+_1 = {p}_a$ and $\hat {p}^+_{n+1} = {p}_b$. It is now clear that there can be at most $n$ roots ${p}_i$.

Now define
\begin{equation}
\tilde f^+_i\left({p} - \Delta {p} + v f^+({p}) \right) =
\cases{
 f^+({p}) & if $\hat {p}^+_i \le {p} \le \hat {p}^+_{i+1}$, \\
 0& otherwise,
}
\end{equation}
which are the monotonic segments with `zero-padding'. One can see that
$\tilde f^+_i\left(\tilde {p}^+({p})\right) = f^+({p})$. Hence, at the roots ${p}_i$ one gets
$f^+({p}_i) = \tilde f^+_i\left(\tilde {p}^+({p}_i)\right) = \tilde f^+_i({p}_0)$. This yields the final result
\begin{equation} \label{EQ_RET}
\tilde f^+({p}_0) = f_{\max} H\left({p}_0 - \tilde {p}^+({p}_b)\right) + \sum_i s_i \tilde f^+_i({p}_0).
\end{equation}

One can understand this result intuitively as follows: Assume that for any given point ${p}_0$ the MLC leaf crosses that point ${p}_0$ twice, once, in segment $n-1$, in the $-{p}$ direction, and then, in
segment $n$, in the $+{p}$ direction. The total fluence is then simply $\tilde
f^+_n({p}_0) - \tilde f^+_{n-1}({p}_0)$, because this is how long point ${p}_0$ is
irradiated. If the leaf crosses the point multiple times, the total fluence is
exactly as given by equation (\ref{EQ_RET}) above.

In words, equation (\ref{EQ_RET}) means that the fluence
delivered is the sum of compressed segments from the original (no motion)
fluence profile, and the negative terms therein ensure a contiguous transition
between those segments. For $f^-({p})$ one gets the result simply by replacing all `$+$' in the exponent with `$-$'. The total overall fluence is $\tilde
f({p}_0) = \tilde f^+({p}_0) - \tilde f^-({p}_0)$.

\end{appendix}

\clearpage

\section*{Acknowledgment}
I would like to thank Drs Steve Webb and Dualta McQuaid from the Royal Marsden Hospital in Sutton, UK, for their thoughtful feedback and constructive
suggestions on the manuscript. In particular,  they worked through the appendix and helped to make it more comprehensible. I would also like to thank my colleague Dr. David Craft from the Massachusetts General Hospital for his thoughtful comments.

This work was supported in part by grants R01-CA118200 and
R01-CA103904 from the National Cancer Institute of the USA.

\section*{References}


\end{document}